\documentclass[aps,prb,superscriptaddress,showpacs,twocolumn,a4paper]{revtex4}

\usepackage{amsmath}
\usepackage{amsfonts}
\usepackage{bm}
\usepackage{graphicx}
\usepackage{color}
\usepackage{hyperref}

\newcommand {\Sm}{\mbox{${\mathbf{\cal{S}}}$}}
\newcommand {\si}{\mbox{\boldmath$\sigma$}}

\begin{document}

\title{Fano effect and nonuniversal phase lapses in mesoscopic
regime of transport}

\author{E.\ R.\ Racec}
\email{roxana@physik.tu-cottbus.de}
\affiliation{Technische Universit\"at Cottbus, Fakult\"at 1, Postfach 101344,
             03013 Cottbus, Germany }
\affiliation{University of Bucharest, Faculty of Physics, PO Box MG-11,
             077125 Bucharest Magurele, Romania}

\begin{abstract}

The transmission probability and phase through a few-electron quantum dot
are studied within a resonance theory for the strong coupling regime
to the conducting leads. We find that the interaction between 
overlapping resonances leads to their separation in the complex 
energy plane and to the  decoherent dephasing of the resonant modes. 
The appearance of the Fano effect is conditioned not only by the induced 
dephasing generally associated with a phase lapse, but also by a favorable 
parity of the resonant modes. 
To identify the contribution of each 
resonance to the transmission, 
we propose, in the case of overlapping resonances,
a decomposition of the
scattering matrix in resonant terms and a background. The resonant 
contributions are approximated by Fano lines with complex asymmetry 
parameters and the background by a constant.
The conductance and the transmission phase calculated 
within this approach reproduce the generic features of the 
experimental data.
\end{abstract}


\maketitle
\section{Introduction}

The Fano effect\cite{fano,fano_book,flach10}, responsible for the 
asymmetric peaks
in the absorption or transmission spectra through quantum systems,
is nowadays one of the most studied phenomena.
This effect was experimentally observed 
for a large variety of quantum systems,
in the neutron scattering\cite{neutron_scattering},
atomic photoionization\cite{fano,fano_book},
Raman scattering\cite{raman_scattering},
optical absorption\cite{optical_absorption}
and, recently, in transport through semiconductor 
quantum dots\cite{goeres,kobayashi02,kobayashi03}. 
The universal presence of the Fano effect
attests the fact that it directly follows from the principles of the 
quantum mechanics. 
According to these principles,
the measurements done on a quantum system necessarily perturb it.
There exists a coupling between the quantum system and  
environment and this coupling causes an indirect
transfer, i.e. by dint of a continuum of states,
between two eigenstates of the 
system.
This process accompanies the direct transition between states
and the two pathways - the direct and the indirect ones -
interfere. This is essentially the mechanism that undergoes the Fano effect.
In a recent study\cite{kroner08} even
the nonlinear Fano effect has been observed in experiments
done on semiconductor quantum dots.
In the nonlinear regime, the direct transition between two eigenstates 
of the quantum system saturates and, after that, in the presence of a 
continuum of states, the indirect transitions become more and more important
leading to a pronounced asymmetry of the line shape in the 
absorption cross-section.

The notable technological progress in the last decade has enabled 
the experimental observation of the Fano effect in mesoscopic systems
and an interesting domain for the theoretical studies  has been opened.
There are two categories of experiments dedicated to this issue.
The first one was  proposed by G\"ores et al.\cite{goeres} in 2000. A quantum
dot strongly coupled to the contacts was defined, by means of top
gates, in a two dimensional electron gas (2DEG) and broad and  asymmetric
profiles were observed in conductance.
They are the signature of the Fano effect
but in this case one can not define two spatially separated interfering 
pathways. The complex asymmetry parameter necessary for fitting
the experimental data reflects the open character of the system,
i.e. a quantum system without time reversal symmetry\cite{levinson02}.
In 2002
Kobayashi et al.\cite{kobayashi02} have proposed a
system in which the two interfering pathways are spatially separated. 
A quantum dot with well separated energy levels is placed
in one arm of an Aharonov-Bohm ring
and the other arm is unperturbed. The electrons travel from the source 
to the drain contact along the two pathways 
and the line shapes of the transmission peaks become broad
and asymmetric\cite{kobayashi02,kobayashi03,ando04}, 
certifying the presence of the Fano effect.
These two experiments were followed by many experimental\cite{kobayashi02,
kobayashi03,kobayashi04,kobayashi05,kalish05,kroner08} 
and theoretical\cite{noeckel94,noeckel95,brouwer01,roxana01,racec02,levinson02,
rotter03,ando04,johnson04,moldoveanu05,satanin05,mendoza08} 
studies dedicated to the 
Fano effect in mesoscopic systems.

For understanding the interference process in 
mesoscopic systems 
some considerations about the transport
mechanism through these systems are needed.
In transmission experiments
an electron is transfered from a state at the Fermi energy
in the source contact to a  state at the Fermi energy
in the drain contact through a quantum system, usually a quantum dot. 
If the quantum system 
is separated from the contacts by high barriers 
the electron has a nonzero probability to tunnel
the barriers and reaches the drain contact. This probability 
is significant if an energy level 
of the quantum system  matches the Fermi energy. In this case 
a resonant tunneling
takes place and the total transmission shows a thin and quasi-symmetric 
maximum, i.e. a Coulomb blockade peak\cite{ferry09}. 
If the confinement barriers 
are lowered, the energy levels of the quantum dot become broad. 
As long as they do not overlap the mechanism described above works further.
The transmission peaks become also broad and slight asymmetric\cite{roxana01}
and their widths correspond to the
width of the resonances supported by the open quantum system.
The asymmetry of the Fano profile in transmission 
has at the origin the existence 
of many resonances; A narrow resonance with the real energy
around the Fermi energy intermediates the {\it direct} electron transfer 
through the quantum system, 
while the other resonances define
{\it indirect} pathways that lead to interference and, consequently, to the
Fano effect\cite{roxana10}. 
To increase further the asymmetry of the Fano profile and therefore 
to amplify the Fano effect in nanostructures, one needs 
overlapping resonances. 
They are supported by scattering potentials
that couple the conducting channels due to their nonseparable character.
In Ref. \onlinecite{roxana10} we showed that
the channel mixing yields pairs of resonances 
very close in energy, but with different widths. 
They are ideal for interference and are responsible for
strong asymmetric peaks and dips in conductance
through a quantum dot. 
A competing process that increases the separation of the resonance
energies is the Coulomb interaction.
In Ref. \onlinecite{karrasch07_2} is reported that
the renormalization effects induced by the local Coulomb interaction
lead to narrow and broad resonant levels that overlap. Each time a 
narrow level crosses the Fermi energy and the broad level,
a Fano-type antiresonance accompanied by a phase lapse 
occurs\cite{karrasch07}. Thus, not only the line shape of the
transmission probability through quantum systems is a fingerprint 
of the Fano effect, but also the phase evolution.

For a systematic analysis of the Fano effect in open quantum systems,
we investigate in this paper a quantum dot strongly coupled 
to the conducting leads via quantum point contacts 
and compute the transmission probabilities and phases through the system.
Similar to the effective one-dimensional
case\cite{roxana01}, the peaks and dips in the
total transmission are approximated as Fano functions with 
a complex asymmetry parameter. Each peak and dip can be associated with 
a resonance or a set of overlapping resonances. In the frame
of our resonance theory\cite{roxana01,roxana10}, we give a method 
to compute the Fano asymmetry parameters 
for an arbitrary potential landscape within the quantum system.
Based on the discussion in Ref. \onlinecite{oreg07},
we analyze further the relation between the asymmetry parameter 
and the phase evolution of the electrons transfered to the quantum dot.
The phase of the transmission probability through open quantum dots
is still an open problem in the mesoscopic physics,
intensively studied experimentally\cite{yacoby95,schuster97,
kalish05,katsumoto07} 
and theoretically \cite{hackenbroich96,oreg97,lee99,yeyati00,oreg02,gefen06,
karrasch07,karrasch07_2,oreg07,bertoni07,mueller09,oreg09,tolea10}
in the last years. In contrast to the universal behavior, specific for
high populated quantum dots,  for the few-electron
quantum dot studied here, we show, as expected, mesoscopic features 
of the transmission phases.

\section{The model}
\label{model}

\begin{figure}[htb]
\begin{center}
\includegraphics[width=3.25in]{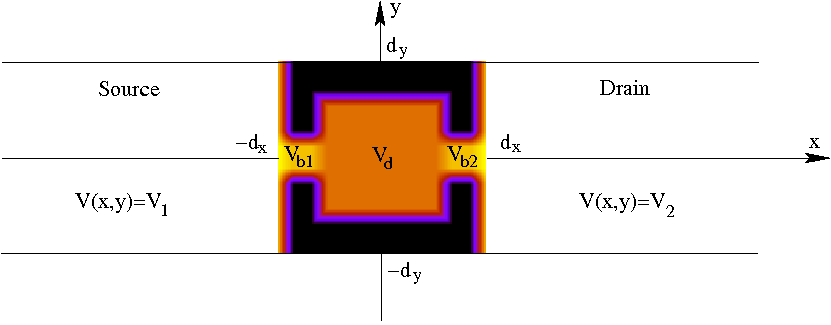}
\caption{
The geometry of a quantum dot embedded inside a quantum wire:
$2d_x=2d_y=175$ nm, the barrier width is $d_b \simeq 35$ nm,
and the point contact regions are about 35 nm $\times$ 35 nm.
Potential energy $V(x,y)$ in the quantum wire 
(dark corresponds to high values and bright to low ones): 
$V_{b0}=100$ meV, $V_{b1}=V_{b2}=2.5$ meV and $V_1 = V_2 =0$.
The potential energy felt by the electrons inside the dot is $V_d$.
At each interface between two domains the potential
energy varies linearly within a distance of 10 nm.
Outside the quantum wire there is a hard wall potential.
The density of 2DEG inside which the quantum dot is defined 
is $N_S=8.1 \times 10^{11}$ cm$^{-2}$
and the Fermi energy is $E_F=29,6$ meV.
}
\label{s_pot_2}
\end{center}
\end{figure}

The transmission phenomena through  weakly confined quantum dots
\cite{goeres,kalish05,goldhaber-gordon,kobayashi02,kobayashi03}
are still a puzzling problem
because the limit of the strong coupling is quite similar to a vanishing
coupling\cite{levinson02}.
Trying to decode this puzzle, we have proposed in Ref. \onlinecite{roxana10}
a model for the scattering in a nonseparable, low confining potential
inside a quantum wire tailored in  2DEG. 
While the positions of the barriers
inside the quantum wire, Fig. \ref{s_pot_2}, are
given by the top gates (see SEM micrograph of the
device in Refs. \onlinecite{goeres} and \onlinecite{kalish05}), 
their heights are the only free parameters in our model.
The two point contacts shown in Fig. \ref{s_pot_2}
ensure  the nonseparable character of the
scattering potential.

The confinement potential of the 
dot and especially the quantum well inside which the electrons are localized
are shallow \cite{kalish05}. The quantum dot created in this way is
quite strongly coupled to the contacts and the open character
becomes dominant determining the transport properties 
through the quantum system.
In the strong coupling regime\cite{goeres,kalish05,goldhaber-gordon,
kobayashi02,kobayashi03},  the two quantum point contacts
act as deep quantum wells ($V_{b1} = V_{b2} \ll V_{b0}$) 
in the lateral direction.

\subsection{Scattering matrix}
\label{scatt_mat}

In Ref. \onlinecite{roxana10} we provide 
a description in detail of the scattering process
in a 2D nonseparable potential in terms of resonances.
The 2D Schr\"odinger equation for the localized scattering potential
$V(x,y)$ is solved directly using the scattering theory and the 
R-matrix formalism.

The scattering functions, 
the most appropriate solutions of the 2D Schr\"odinger equation
in the case of a localized scatterer,
are given in terms of the generalized scattering matrix
$\Sm$.
Outside the scattering region, i.e. the dot region,
they have the expressions \cite{roxana10}
\begin{widetext}
\begin{equation}
\psi_n^{(s)}(E;x,y)
= \frac{\theta[N_s(E)-n]}{\sqrt{2 \pi}}
  \begin{cases}
     \delta_{s1} \exp{[ i \; k_{1n} \; (x+d_x)]} \; \phi_n(y)
    +\sum_{n'=1}^{\infty}\limits \Sm^T_{sn,1n'}(E)
                          \exp{[-i \; k_{1n'} \; (x+d_x)]} \; \phi_{n'}(y),
    & x \le -d_x \cr
     \delta_{s2} \exp{[-i \; k_{2n} \; (x-d_x)]} \; \phi_n(y)
    +\sum_{n'=1}^{\infty}\limits \Sm^T_{sn,2n'}(E)
                          \exp{[ i \; k_{2n'} \; (x-d_x)]} \; \phi_{n'}(y),
    & x \ge d_x
  \end{cases}
\label{psi-outside2}
\end{equation}
\end{widetext}
with $s=1$ for the source contact and 
$s=2$ for the drain contact. 
The eigenfunctions
\begin{equation}
\phi_n(y) = \frac{1}{\sqrt{d_y}} 
            \sin \left[ \frac{n \pi}{2d_y}(x+d_y) \right],
\label{phi_n}
\end{equation}
associated with the
lateral problem in contacts
satisfy Dirichlet boundary conditions
imposed by the  isolated quantum wire
and their quantum number $n \ge 1$ defines the scattering 
energy channels\cite{roxana10}. 
The wave vectors have the expression 
$k_{sn}(E) = k_0 \sqrt{(E-E_{\perp n}-V_s)/u_0}$
with $k_0=\pi/2d_x$, $u_0 = \hbar^2 k_0^2/2 m^*$
and $E_{\perp n}=n^2 u_0 d_x^2/d_y^2$.
The Heaviside function $\theta$ in Eq. (\ref{psi-outside2})
cuts the solutions without physical meaning;
The number of conducting channels for the energy $E$, $N_s(E)$,
is the greatest value of $n$ for which $k_{sn}$ is real
and allows for incoming plane waves, $s=1,2$. 
The rest of channels, $n > N_s(E)$, are nonconducting
or evanescent ones\cite{racec09}.
For the structure tailored in a 2DEG
the potential well has no attractive character and
the nonconducting channels become important for the transport
only for a very low electron density in the contacts. 

The current scattering matrix ${\tilde{\Sm}}$,
simply called scattering matrix,
is related to $\Sm$ by the definition 
\begin{equation}
{\tilde{\Sm}} = \mathbf{K}^{1/2} \mathbf{\Theta} \Sm \mathbf{K}^{-1/2},
\label{S_def}
\end{equation}
where the wave vector matrix $\mathbf{K}$  is a diagonal one,
$\mathbf{K}_{sn,s'n'}(E)= k_{sn}(E) \, \delta_{nn'} \delta_{ss'} /k_0$.
Due to the matrix $\Theta$ with $\mathbf{\Theta}_{sn,s'n'}(E)
= \theta[N_s(E)-n] \, \delta_{ss'} \, \delta_{nn'}$
only the matrix elements of ${\tilde{\Sm}}$ corresponding
to conducting channels are nonzero.
So that, for a given confining potential $V(x,y)$ 
of the open quantum dot,
the transport properties through the dot are determined as a function
of the scattering matrix ${\tilde{\Sm}}$.

In the frame of the R-matrix formalism\cite{roxana10}
the current scattering matrix is given by
\begin{equation}
{\tilde{\Sm}} = \mathbf{\Theta}
                \left[ \mathbf{1} - 2 (\mathbf{1} + i \mathbf{\Omega})^{-1}
                \right]
                \mathbf{\Theta},
\label{Stilde2}
\end{equation}
with the symmetrical infinite matrix 
\begin{equation}
\mathbf{\Omega}(E) = u_0 \sum_{l=1}^{\infty}
                      \frac{\vec{\alpha}_l \, \vec{\alpha}^T_l}
                           {E-E_l},
\label{omega}
\end{equation}
and the column vector 
\begin{equation}
(\vec{\alpha}_l)_{sn} =  \frac{k_{sn}}{k_0\sqrt{k_0}} 
                         \int_{-d_y}^{d_y} dy \; \chi_l[(-1)^{s} d_x,y]
                                               \phi_n(y)
\label{alpha}
\end{equation}
expressed in terms of the Wigner-Eisenbud functions $\chi_l(x,y)$ and 
energies $E_l$, $l \ge 1$.
They are solutions of the Wigner-Eisenbud problem
\begin{equation}
\left[-\frac{\hbar^2}{2 m^*} 
\left( \frac{\partial^2}{\partial x^2}
      +\frac{\partial^2}{\partial y^2}
\right) 
      +V(x,y)
\right] \; \chi_l(x,y)
= E_l \; \chi_l(x,y),
\label{WEp}
\end{equation}
defined within the scattering area, $-d_x \le x \le d_x$
and $-d_y < y < d_y$. The functions $\chi_l$
satisfy homogeneous Neumann boundary
conditions at the interfaces between dot and contacts,
$\left. \partial\chi_l/\partial x \right|_{x=\pm d_x}=0$,
and Dirichlet boundary conditions on the surfaces perpendicular 
to the transport direction, $\chi_l(x,\pm d_y)=0$.
The Wigner-Eisenbud functions and energies are real as long as no magnetic 
field acts on the scattering area.

The current scattering matrix  gives directly
the reflection and transmission probabilities through the
quantum dot. The probability of an electron incident from the 
contact $s=1,2$ on the channel $n$
to be transmitted into the contact $s' \ne s$ on the channel $n'$
is defined as
\begin{equation}
T_{nn'}(E)  = \left| \si_{n'n}(E) \right|^2
            = \left| \si_{nn'}(E) \right|^2,
\label{Tnnp_1}
\end{equation}
where $\si$ is the part of $\tilde{\Sm}$
that contains
the transmission amplitudes, 
\begin{equation}
\si_{nn'}(E)=\tilde{\Sm}_{2n',1n}(E),
\label{sigma}
\end{equation}
$n,n' \ge 1$.
Per construction the matrix $\si$ is symmetric.
For the nonconducting  channels the transmission probabilities are zero,
$T_{nn'}=0$ for $n > N_1(E)$ or $n' > N_2(E)$.  
The sum of all transmission coefficients defines 
the total transmission through the quantum system
\begin{equation}
T(E)= \sum_{n=1}^{N_1(E)} \sum_{n'=1}^{N_2(E)} T_{nn'}(E).
\label{TT}
\end{equation}

The restriction of the $\tilde{\Sm}$-matrix to the conducting channels,
${\bf S}_{sn,s'n'}=\tilde{\Sm}_{sn,s'n'}$, $s,s'=1,2$, $n=\overline{1,N_s}$,
$n'=\overline{1,N_{s'}}$, 
is the well
known unitary current transmission matrix \cite{wulf98,roxana01,racec09}
commonly used in the
Landauer-B\"uttiker formalism,
$ \mathbf{S} \mathbf{S}^\dagger =\mathbf{S}^\dagger \mathbf{S}
                                ={\mbox{1}}$,
where $\mathbf{S}^\dagger$ denotes the adjoint matrix.

Based on the relation (\ref{Stilde2}) the scattering matrix  
is immediately constructed using the Wigner-Eisenbud 
functions and energies. The R-matrix method does not require a 
numerical solution of an eigenvalue
problem for every energy and this peculiarity
leads to a high numerical efficiency of the method.

\begin{figure}[t]
\begin{center}
(a)
\noindent\includegraphics*[width=2.75in]{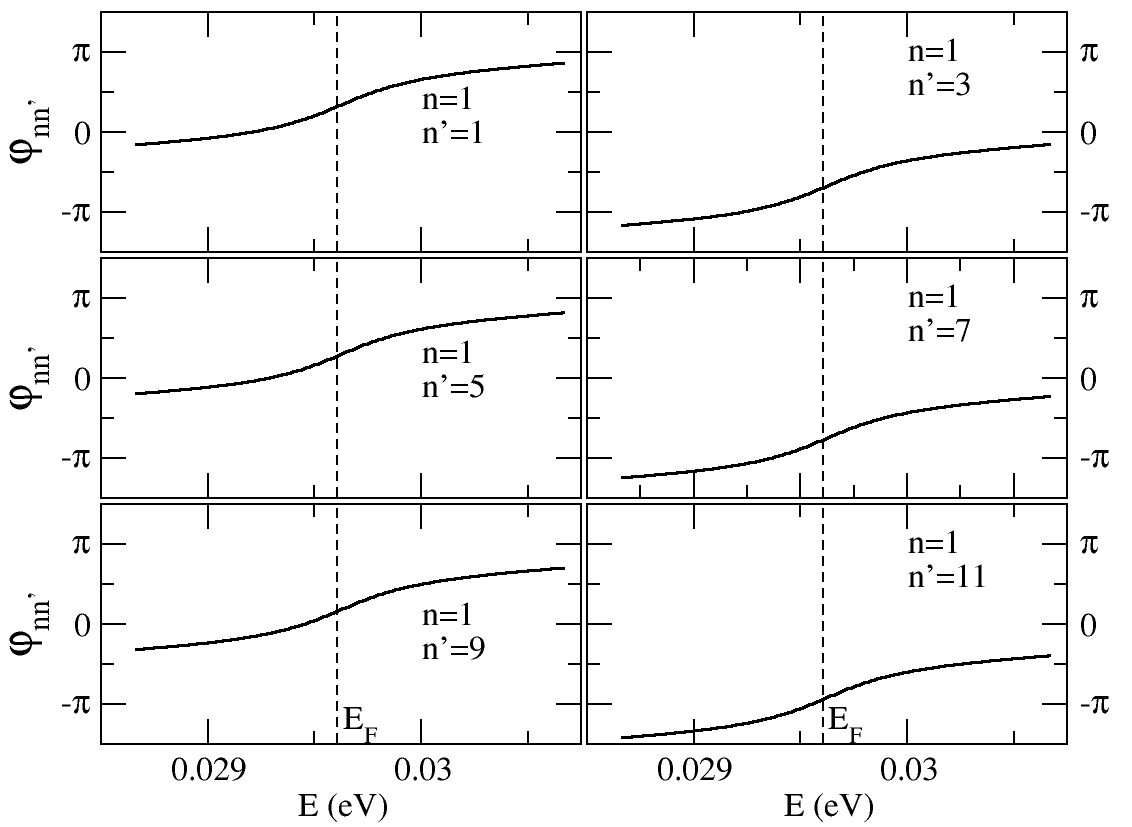}
(b)
\noindent\includegraphics*[width=2.75in]{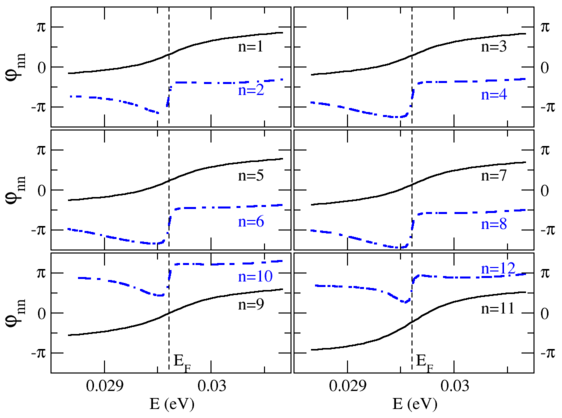}
\end{center}
\caption{(Color online)
Transmission phases between scattering channels with
the same parity (solid black lines for odd channels 
and dot-dashed blue lines for the even ones):
(a) $n=1$, $n'=\overline{1,N_F}$ and (b) $n=n'=\overline {1,N_F}$.
The potential energy in the dot region is $V_d=V_0^{(1,2)}$ 
for which $E_0^{(1,2)} \simeq E_F$.}
\label{TT_fi}
\end{figure}

Besides the transmission probability through an open quantum system, 
Eq. (\ref{Tnnp_1}), the phase of the transmission amplitude\cite{kalish05},
\begin{equation}
\varphi_{nn'}(E) = \arg[\si_{nn'}(E)],
\label{phase}
\end{equation}
provides complementary information about the transport properties.
In Fig. \ref{TT_fi} some transmission phases $\varphi_{nn'}$ 
are plotted as a function of energy around $E_F$,
were $N_s(E_F)=N_F=12$, $s=1,2$ for the considered quantum system.
The potential energy in the dot region $V_d$ 
has the value $V_0^{(1,2)}$ for
which two overlapping resonances $E_0^{(1,2)}$ and 
$E_0^{(2,1)}$ lie around the Fermi energy\cite{roxana10}.
The phases associated with the odd scattering channels, 
$\varphi_{nn'}$ with $n,n'=1,3,...,11$,
change slowly by $\pi$ in an energy interval comparable to the width
of the resonance $(2,1)$, $\Gamma^{(2,1)}= 6 \cdot 10^{-4}$ eV,
while for the even channels, the transmission phases show
a total different evolution through the resonance $(1,2)$.
They decrease slowly by approximatively $0.9 \pi$ 
and increase after that abruptly to the initial value, similar to a 
phase lapse\cite{kalish05,oreg97,karrasch07,oreg07}  .
For symmetry reasons the transmissions
between odd and even scattering channels are forbidden.

Two conclusions are drawn from the plots in Fig. \ref{TT_fi}:
i) the channel parity given by the parity of the function $\phi_n(y)$,
Eq. (\ref{phi_n}), plays a decisive role in the transmission 
processes through an open quantum system
and ii) the transmission phases between channels with
the same parity show similar energy dependence 
in the energy domain of a resonance.
So that, for a qualitative characterization of 
the phase variation through a resonance it is enough to analyze 
the phases $\varphi_{11}$ and $\varphi_{22}$. 
The measured transmission phase through the quantum system,
a linear combination of all phases $\varphi_{nn'}$,
should be approximatively obtained as a linear combination of 
$\varphi_{11}$ and $\varphi_{22}$.

\subsection{Resonances}
\label{resonances}

The transport through an open quantum dot strongly 
coupled to the conducting leads is dominated by resonances.
When the quantum system becomes open, the corresponding 
Hamilton operator becomes non-Hermitian \cite{mueller09,rotter09}
and the eigenstates $E_\lambda$  of the closed dot
become resonances with the complex energies $\bar{E}_{0\lambda} 
= E_{0\lambda} - i \Gamma_\lambda/2$, $\lambda \ge 1$.
They are  simple poles of the current 
scattering matrix ${\tilde{\Sm}}$,  solutions of the equation
$
\det[{\bf 1} + i {\mathbf \Omega}]=0.
$ As shown in Ref. \onlinecite{roxana10},
the R-matrix formalism allows 
for a decomposition of ${\tilde{\Sm}}$ around $E_\lambda$ 
in which the resonant and 
the background parts are
separated,
\begin{equation}
{\tilde{\Sm}}(E) = 2 i u_0 \frac{\mathbf{\Theta} \, \vec{\beta}_\lambda
                                 \,
                                 \vec{\beta}^T_\lambda \, \mathbf{\Theta}}
                                {E - E_\lambda -\bar{\cal{E}}_\lambda(E)}
                   +{\tilde{\Sm}}_\lambda(E),
\label{Stilde3}
\end{equation}
where
$\vec{\beta}_\lambda(E) = (\mathbf{1} + i \mathbf{\Omega}_\lambda)^{-1}
                         \vec{\alpha}_\lambda$
is an infinite column vector that characterizes the resonance $\lambda$,
$\bar{\cal{E}}_\lambda(E)=-i\vec{\beta}^T_\lambda \cdot \vec{\alpha}_\lambda$
is a complex function which ensures the analyticity of the current
scattering matrix
for every real energy $E$ and
the matrix $\mathbf{\Omega}_\lambda$
is obtained from $\mathbf{\Omega}$ taking out the $\lambda$ term,
$\mathbf{\Omega}_\lambda
=\mathbf{\Omega}-u_0 \frac{\vec{\alpha}_{\lambda} \,
                           \vec{\alpha}^T_{\lambda}}
                           {E-E_{\lambda}}$.
The background matrix
\begin{equation}
{\tilde{\Sm}}_\lambda(E) = \mathbf{\Theta}
                       \left[ \mathbf{1}
                             -2 (\mathbf{1} + i \mathbf{\Omega}_\lambda)^{-1}
                       \right]
                       \mathbf{\Theta}
\label{Sbg}
\end{equation}
has a similar expression to the scattering matrix $\tilde{\Sm}$, 
Eq. (\ref{Stilde2}), and gives rise to a further decomposition 
in a resonant and a background term if a second resonance exists
around the Fermi energy. 

The position of the resonance
$\bar{E}_{0 \lambda}$ in the complex energy plane is determined as a
solution of the equation
\begin{equation}
 \bar{E}_{0 \lambda} - E_\lambda 
-\bar{\cal{E}}_\lambda(\bar{E}_{0 \lambda}) = 0.
\label{eq-poles}
\end{equation}
There exists a resonance energy associated with every Wigner-Eisenbud energy
and the above equation is solved numerically  
using an iterative procedure starting with $\bar{E}=E_\lambda$.
This method gives all resonances supported by the open 
quantum system, even the very narrow ones
associated with quasi-bound states mainly localized within the dot region
or the very broad ones associated with modes mainly localized
in the point contact regions, i.e. {\it aperture modes}\cite{roxana10}.
Based on the analogy between the scattering 
functions 
at the resonance energy, $|\psi_n^{(s)}(E_{0\lambda};x,y)|^2$, and
the eigenfunctions of the corresponding closed quantum dot, 
we associate a pair of quantum numbers $(n_x,n_y)$ with each
resonance $\lambda$; $n_x$ denotes the number of the maxima
of $|\psi_n^{(s)}(E_{0\lambda};x,y)|^2$ 
in the $x$-direction 
and $n_y$ the number of the maxima in the $y$-direction and  $(x,y)$ are  
points in the dot region. 
As discussed in Ref. \onlinecite{roxana10},
the index $\lambda$ of a
resonance energy is only technically relevant,
but has no physical meaning.

\subsection{Fano Approximation}
\label{fano}

Based on the decomposition, Eq. (\ref{Stilde3}), of the $\tilde{\Sm}$
matrix, the transmission 
probability between the channel $n$ in the source contact
and the channel $n'$ in the drain contact can be written
as a sum of a resonant contribution and a background,
\begin{equation}
T_{nn'}(E) = T_{res}^{(nn')}(E) + T_{bg}^{(nn')}(E)
\label{Tnnp}
\end{equation}
for energies around a given resonance energy $E_{0 \lambda}$.
The first contribution to the transmission probability
contains a resonant term singular at
$E=\bar{E}_{0 \lambda}$,
\begin{equation}
T_{res}^{(nn')}(E) 
= \left| \frac{2 i {\cal{Z}}_{1 \lambda}^{(nn')}(E)}
              {E - E_\lambda -\bar{\cal{E}}_\lambda(E)}
        +{\cal{Z}}_{2 \lambda}^{(nn')}(E)
  \right|^2
 -\left| {\cal{Z}}_{2 \lambda}^{(nn')}(E) \right|^2
\label{Tnnpres}
\end{equation}
and the slowly varying functions of energy in this term are defined as
\begin{equation}
{\cal{Z}}_{1 \lambda}^{(nn')}(E) 
= \left| (\vec{\beta}_\lambda)_{1n} \right|
  \left| (\vec{\beta}_\lambda)_{2n'} \right|
\label{Z1nnp}
\end{equation}
and 
\begin{equation}
{\cal{Z}}_{2 \lambda}^{(nn')}(E) 
= (\vec{\beta}^*_\lambda)_{1n}
  (\si_\lambda)_{nn'}
  (\vec{\beta}^*_\lambda)_{2n'}
  /{\cal{Z}}_{1 \lambda}^{(nn')}(E),
\label{Z2nnp}
\end{equation}
where $\si_\lambda$ is the part of $\tilde{\Sm}_\lambda$
that contains the transmission amplitudes,
$(\si_\lambda)_{nn'}(E)=(\tilde{\Sm}_\lambda)_{2n',1n}(E)$;
$\vec{\beta}^*_\lambda$ denotes the complex conjugate of 
$\vec{\beta}_\lambda$.
The background term
\begin{equation}
T_{bg}^{(nn')}(E) =  \left| (\si_\lambda)_{nn'}(E) \right|^2
\label{Tbgnnp}
\end{equation}
is also a slowly varying function of energy around $E_{0 \lambda}$,
if a second resonance does not exist in the energy domain of the
resonance $\lambda$.

As shown in Appendix \ref{laurent},
the contribution of the resonance $\bar{E}_{0 \lambda}$ to the 
transmission probability $T_{nn'}$
can be approximated as a Fano line
\begin{equation}
T^{(nn')}_{res}(E)
\simeq
T_{1 nn'}
\left| \frac{1}{e_\lambda +i} + \frac{1}{q_{Fnn'}} \right|^2
-T_{2 nn'},
\label{TnnpF}
\end{equation}
with a complex asymmetry parameter $q_{Fnn'}$, where
$e_\lambda = 2(E-E_{0 \lambda})/\Gamma_\lambda$.
The constants $T_{1 nn'}$, $T_{2 nn'}$ and $q_{Fnn'}$,
computed using Eqs. (\ref{qFF}), (\ref{F1F}) 
and (\ref{F2F}) respectively,
depend actually on $\lambda$, 
but this index was omitted for simplicity.
For an isolated resonance
the background contribution to the conductance is approximated as a
constant, $T_{bg}^{(nn')}(E) \simeq T_{bg}^{(nn')}(E_{0 \lambda})$.

The Fano line is defined as the absolute value of an 
energy dependent complex function and,
as presented in Appendix \ref{fano_fct_c},
one can associate a phase
\begin{equation}
\varphi_{Fnn'} = \arg \left(\frac{1}{e_\lambda +i} + \frac{1}{q_{Fnn'}}
                      \right),
\label{Fano_phase}
\end{equation}
with each Fano line.
The Fano phase, $\varphi_{Fnn'}$ is an energy dependent function 
that provides a good approximation for 
the phase of the transmission amplitude
around a resonance, 
\begin{equation}
\varphi_{nn'} \simeq \varphi_{Fnn'} + \varphi^{(0)}_{nn'}.
\label{fi_a}
\end{equation}
The constant phase $\varphi^{(0)}_{nn'}$ can not be determined in the
limits of the considered approach, because the Fano line gives 
only a satisfactory description on the peak, without the information about 
the background.
In Appendix \ref{fano_fct_c} we have analyzed  
the Fano lines and the associated Fano phases
for different values of the complex asymmetry parameter $q_F=(q_r,q_i)$.
The phase evolution through a peak, Fig. \ref{fano_g}, 
shows typical mesoscopic features as observed experimentally 
by Avinun-Kalish et al., Ref. \onlinecite{kalish05}.
For an asymmetry parameter $q_F$ with $q_i<-1$
the phase increases monotonically by $2\pi$, while for $q_i>-1$
there is no global variation of the phase through the Fano peak;
The phase has a maximum and a minimum in the peak region and reaches
the initial value far away from the peak. In the case 
$|q_r| \gg |q_i|$ typical phase lapses
of $\pi$ occur, while for $|q_i| > |q_r|$
the maxima and minima are approximatively equal in amplitude
and generally smaller than $\pi$. This last type of phase 
variation through a resonance is also reported by Avinun-Kalish et al., 
see Figs.  4(b) and 5(b) in Ref. \onlinecite{kalish05},
and gives a direct experimental proof of the complex nature
of the Fano asymmetry parameter. 
In the limit $q_i \rightarrow 0$, i.e. neglecting the breaking 
of the time reversal symmetry in open systems, the existence 
of the phase lapses can be explained\cite{oreg07}, 
but the whole variety of profiles
in the phase evolution observed experimentally in the mesoscopic regime
can not be obtained. 

The analysis of the complex Fano function in Appendix \ref{fano_fct_c} 
proves that the Fano
phase is more sensitive to the value of the asymmetry parameter 
than the  well-known real Fano function. 
If the phase evolution
in the transmission peak region is known, the domain in the complex
plane where the Fano asymmetry parameter takes values 
is considerably restricted and it can be 
more precisely computed as a fit parameter for the
transmission probability. The Fano parameter contains 
information about the processes that accompany the scattering 
through the open quantum dot and a possible exact value of it
leads to a better characterization of the transport properties through 
the system.
A strong asymmetry of the Fano line ($|1/q_F| > 1$) indicates a strong 
interaction between overlapping resonances due to  channel 
mixing\cite{roxana10}.
The quantum interference becomes dominant in this type of open
systems that can not be anymore described
in the frame of an effective one dimensional model\cite{roxana10}.
Supplementary, the imaginary part of the Fano asymmetry parameter
reflects the decoherence processes
induced in an open quantum system by the interaction 
with the conducting leads\cite{brouwer01,srotter10}. 
Even for systems for which the dissipation can be neglected,
the coupling of the quantum system to the environment  
allows for a decoherent dephasing
in the limits of an unitary evolution\cite{srotter10}.

\section{Conductance and transmission phases}
\label{conductance}

\begin{figure*}[t]
\begin{center}
\includegraphics[width=5in]{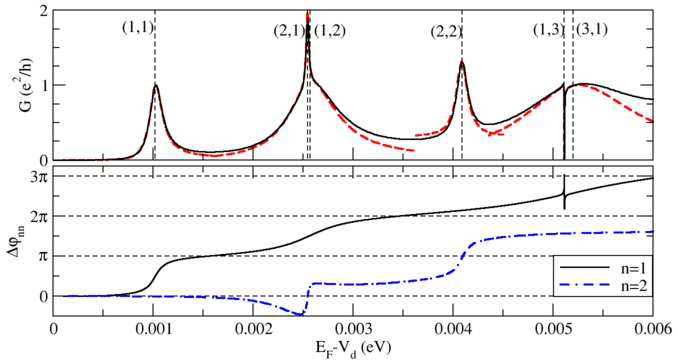} 
\caption{
Upper part: Conductance (solid black line) and its Fano approximation
(dashed red line) on each peak as a function of the 
potential energy in the dot region.
Lower part: Phases of the transmission amplitudes between the 
scattering channels (1,n) and (2,n) with $n=1$ (black solid line)
and $n=2$ (dot-dashed blue line). The phases for $V_d=E_F$ 
are set to $0$ in order to make evidently the phase variations of $\pi$.
}
\label{GG1}
\end{center}
\end{figure*}

\begin{figure*}[t]
\begin{center}
\includegraphics[width=5in]{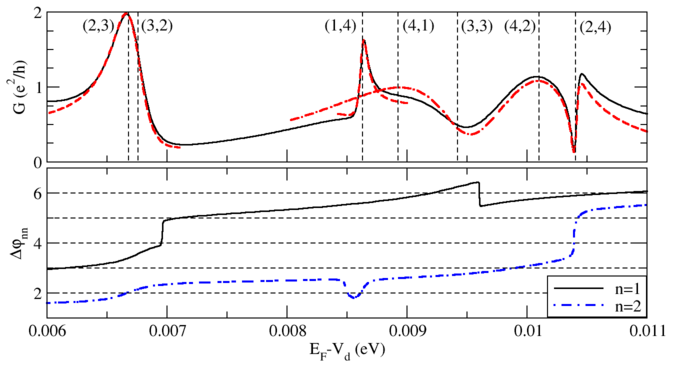}
\caption{
Upper part: Conductance (solid black line) and its Fano approximation
(dashed red line) on each peak as a function of the
potential energy in the dot region.
Lower part: Phases of the transmission amplitudes between the
scattering channels (1,n) and (2,n) with $n=1$ (solid black line)
and $n=2$ (dot-dashed blue line). The phases for $V_d=E_F$
are set to $0$ in order to make evidently the phase variations of $\pi$.
}
\label{GG2}
\end{center}
\end{figure*}

In the frame of a noninteracting model and for low temperatures
the conductance\cite{roxana10} through an open quantum system
is directly related to the total transmission, Eq. (\ref{TT}),
at the Fermi energy,
\begin{equation}
G(V_d) =\frac{e^2}{h} \,T(E_F;V_d),
\label{G}
\end{equation}
where $V_d$ is the potential energy experienced by the 
electrons within the quantum system.
The transmission coefficients are determined, as
shown in Sec. II, for a fixed value of $V_d$. 
In the case of a quantum dot confined within a very shallow quantum 
well , i.e. for a low occupancy of the dot,  
the conductance, Figs. \ref{GG1} and \ref{GG2}, 
%\ref{GG3}, 
shows typical mesoscopic features: 
asymmetrical peaks and dips whose widths
do not increase monotonically with the distance to the bottom of 
the quantum dot, $E_F-V_d$. 
They are associated with
isolated resonances or with sets of overlapping resonances\cite{roxana10}.
For each such a peak and dip we define the potential energy 
$V_{0\lambda}=V_0^{(n_x,n_y)}$
for which the main resonance of the set
participates to the conductance, i.e. the resonance energy matches 
the Fermi energy, $E_{0 \lambda} =E_0^{(n_x,n_y)} \simeq E_F$. 
The main resonance, denoted by $\lambda$, is
the thinnest one of the set of overlapping resonances at the Fermi energy.
The other resonances of the set are indexed according to 
their widths by $\lambda'$, $\lambda''$ and so on. 

To connect the peaks in the conductance to the resonances  and 
to the scattering phases,  
one needs a functional dependence of the
scattering matrix  on $V_d$ around $V_0$; Here the index $\lambda$
is omitted for simplicity.
A small variation $\delta V=V_d-V_0$ of the potential energy
felt by the electron in the dot region can be approximately
seen as a shift of the potential energy in the whole scattering area
and, in turn, the conductance around $V_0$
becomes\cite{roxana10} 
\begin{widetext}
\begin{equation}
G(V_0+\delta V) \simeq G_{res}(E_F-\delta V; V_0) + G_{bg}(E_F-\delta V; V_0),
\label{Ga}
\end{equation}
where the energy dependent functions $G_{res}$ and $G_{bg}$, given as
\begin{equation}
G_{res}(E) = \frac{e^2}{h} 
         \left[ \left| \frac{2 i {\cal{Z}}_{1 \lambda}(E)}
                            {E - E_\lambda -\bar{\cal{E}}_\lambda(E)}
                      +{\cal{Z}}_{2 \lambda}(E)
                \right|^2
               -\left| {\cal{Z}}_{2 \lambda}(E) \right|^2
         \right],
\label{Gres}
\end{equation}
\end{widetext}
and 
\begin{equation}
G_{bg}(E) = \frac{e^2}{h} \mbox{Tr} [\mathbf{\sigma}_\lambda(E)
                                    \mathbf{\sigma}_\lambda^\dagger(E)],
\label{Gbg}
\end{equation}
respectively, are exact expressions obtained from Eq. (\ref{Stilde3})
without any approximation.
The functions 
\begin{equation}
{\cal{Z}}_{1 \lambda}(E) = \left| \vec{\beta}_{1 \lambda} \right|
                                  \left| \vec{\beta}_{2 \lambda} \right|
\label{Z1}
\end{equation}
and 
\begin{equation}
{\cal{Z}}_{2 \lambda}(E) = \vec{\beta}_{1 \lambda}^\dagger
                           \mathbf{\sigma}_\lambda
                           \vec{\beta}_{2 \lambda}^*
                          /{\cal{Z}}_{1 \lambda}(E)
\label{Z2}
\end{equation}
are generally slowly varying in the energy domain of the resonance $\lambda$; 
$(\vec{\beta}_{1 \lambda})_n = (\vec{\beta}_\lambda)_{1n}$,
$(\vec{\beta}_{2 \lambda})_n = (\vec{\beta}_\lambda)_{2n}$,
and $({\mathbf{\sigma}}_\lambda)_{nn'} =
({\tilde{\mathbf{\cal{S}}}}_\lambda)_{1n,2n'}$, 
$n\le N_1(E_F)$, $n' \le N_2(E_F)$.
Thus, the first contribution to the conductance, $G_{res}$, 
contains a resonant term singular at 
$E=\bar{E}_{0 \lambda}$ and a term, 
${\cal{Z}}_{2 \lambda}(E)$, that describes the coupling of the
resonance $\lambda$ characterized by the vector $\vec{\beta}_\lambda$ 
to the other resonances characterized by the background matrix 
$\si_\lambda$. 
The second contribution to the conductance, $G_{bg}$,
is  given only by the background matrix $\si_\lambda$
and it is approximatively constant in the case of an isolated resonance.

In the presence of the second resonance $\lambda' \ne \lambda$
around the Fermi energy, 
the background contribution 
to the conductance, $G_{bg}$, has an energy
dependence that can not be neglected anymore.
The background term $\tilde{\Sm}_\lambda$ 
in $\tilde{\Sm}$, Eq. (\ref{Stilde3}), can be further decomposed\cite{roxana10} 
into a resonant term corresponding to the resonance $\lambda'$ 
and a second background $\tilde{\Sm}'_\lambda$ .
Consequently,  the term $G_{bg}$ in the expression Eq. (\ref{Gbg}) 
of the conductance
is written as a sum of two contributions,
\begin{widetext}
\begin{equation}
G_{bg}(E_F-\delta V;V_0) = G'_{res}(E_F-\delta V;V_0) 
                          +G'_{bg}(E_F-\delta V;V_0),
\label{Gbg2}
\end{equation}
a resonant one, 
\begin{equation}
G'_{res}(E) = \frac{e^2}{h} 
         \left[ \left| \frac{2 i {\cal{Z}}'_{1 \lambda}(E)}
                            {E - E_{\lambda'} -\bar{\cal{E}}'_\lambda(E)}
                      +{\cal{Z}}'_{2 \lambda}(E)
                \right|^2
               -\left| {\cal{Z}}'_{2 \lambda}(E) \right|^2
         \right],
\label{Gresp}
\end{equation}
\end{widetext}
and a second background
\begin{equation}
G'_{bg}(E) = \frac{e^2}{h} \mbox{Tr} [\mathbf{\sigma}'_\lambda(E)
                                    \mathbf{\sigma}_\lambda^{'\dagger}(E)],
\label{Gbgp}
\end{equation}
slowly varying with the energy if a third resonance does not
exist around $E_F$, where 
$(\si'_\lambda)_{n n'}= (\tilde{\Sm}'_\lambda)_{2n,1n'}$, $n,n' \ge 1$.
The functions ${\cal{Z}}'_{1 \lambda}$,
${\cal{Z}}'_{2 \lambda}$, and $\bar{\cal{E}}'_\lambda$
are obtained from ${\cal{Z}}_{1 \lambda}$, ${\cal{Z}}_{2 \lambda}$, and
$\bar{\cal{E}}_\lambda$ by replacing
$\vec{\alpha}_\lambda$ by $\vec{\alpha}_{\lambda'}$
and $\mathbf{\Omega}_\lambda$ by
$\mathbf{\Omega}'_\lambda = \mathbf{\Omega}_\lambda
                           -u_0 \frac{\vec{\alpha}_{\lambda'} \,
                                      \vec{\alpha}^T_{\lambda'}}
                                     {E-E_{\lambda'}}$.
In Eq. (\ref{Gbg2}) the potential energy $V_0$ denotes $V_{0\lambda}$
corresponding to the thinest resonance of the set.  

The expression Eq. (\ref{Gbgp}) of the resonant contribution to the conductance
allows for a new split up of this term in the case of three
interacting resonances around the Fermi energy. 
The successive decomposition of the scattering matrix and, in turn, of
the conductance, presented in this section is relevant as long as 
it is performed hierarchically, from the thinnest resonance to the 
broadest one and can be used for an arbitrary number of overlapping resonances.

\subsection{Isolated resonances}
\label{is_res}

In the case of an isolated resonance\cite{roxana10}
around the Fermi energy, $\bar{E}_{0 \lambda} \simeq E_F$, 
the contribution $G_{res}$ to the conductance, Eq. (\ref{Gres}), 
is singular at
$E=\bar{E}_{0 \lambda}$ and
yields always a peak mainly localized
in the resonance domain.
Following the method presented in Appendix \ref{laurent},
this contribution can be approximated as a Fano line
with the complex asymmetry parameter $q_{F\lambda}$,
\begin{eqnarray}
G_{res}(E_F-\delta V)
& \simeq &
G_{1 \lambda}
\left| \frac{1}{v_\lambda +i} + \frac{1}{q_{F \lambda}} \right|^2
-G_{2 \lambda} \nonumber \\
& = & G_F(E_F-\delta V),
\label{GF}
\end{eqnarray}
where $v_\lambda = -2 \delta V/\Gamma_\lambda$.
The constants $G_{1 \lambda}$ and $G_{2 \lambda}$
as well as the Fano parameter $q_{F\lambda}$ 
are computed using Eqs. (\ref{F1F}), (\ref{F2F}) and (\ref{qFF}), 
respectively.
The background contribution to the conductance is approximated by a
constant,
\begin{equation}
G_{bg}(E_F-\delta V) \simeq G_{bg}(E_F) = G_0.
\label{G0}
\end{equation} 

According to Eq. (\ref{GF})
the lowest approximation for a resonant peak in conductance 
through an open quantum dot is a  Fano
line with a complex asymmetry parameter.
For huge values of this parameter,
i.e. in the limit $|1/q_{F \lambda}| \ll 1$,
the line shape becomes quasi Breit-Wigner.
From the mathematical point of view, this situation corresponds to a 
very slow variation with the energy of the functions
${\cal{Z}}_{1 \lambda}$ and $\bar{\cal{E}}_\lambda$. 
The condition is fulfilled only for an almost constant rest matrix 
${\tilde{\Sm}}_\lambda$ in the energy domain of the resonance $\lambda$, 
i.e. if and only if the other
resonances are far enough from the considered one.
Supplementary, ${\tilde{\Sm}}_\lambda$ should have small values 
comparatively to the singular term in Eq. (\ref{Stilde3}).
In the presence of a weak interaction with other resonances
$|1/q_{F \lambda}|$ increases 
and the corresponding line in the total transmission through the quantum dot
becomes asymmetric.

\begin{table}
\begin{tabular}[t]{|c|p{2.5cm}|p{2.cm}|p{2.cm}|}
\hline
\hline
Resonance & $q_F$           & $q_{F11}$       & $q_{F22}$ \\
\hline
(1,1)     & $9.399-1.003 \,i$  & $9.411-1.06 \,i$  &    -              \\ 
(1,2)     & $50.45-1.013 \,i$  &     -             & $15.3+17.1 \, i$  \\ 
(2,1)     & $13.38-1.049 \,i$  & $12.33-5.04 \,i$  &    -              \\ 
(2,2)     & $144.7-1.052 \,i$  &       -           & $-0.47-21.1\,i$   \\ 
(1,3)     & $-0.23-1.0002\,i$  & $-0.23-0.98 \,i$  &    -              \\ 
(3,1)     & $11.29-1.248 \,i$  & $9.386-5.93 \,i$  &    -              \\ 
(2,3)     & $-1.22-0.999 \,i$  & $-1.25-1.19 \,i$  &    -              \\ 
(3,2)     & $-67.7-1.196 \,i$  &       -           & $-24.19+18.8 \,i$ \\ 
(1,4)     & $3.144-1.001 \,i$  &       -           & $3.361-0.64  \,i$ \\ 
(4,1)     & $9.78-1.197  \,i$  & $7.44-4.36 \,i$   &    -              \\ 
(3,3)     & $-0.27-0.995 \,i$  & $-0.27-0.99 \,i$  &    -              \\ 
(2,4)     & $0.52-1.003 \,i$   &       -           & $0.5-1.059 \,i$   \\ 
(4,2)     & $-12.01-0.883 \,i$ &       -           & $-5.84-6.12 \,i$  \\ 
\hline
\hline
\end{tabular}
\caption{Asymmetry parameters of the Fano profiles in conductance and 
in the transmission probabilities $T_{11}$ and $T_{22}$ for 
the indicated resonances. 
}
\label{tabel}
\end{table}

The proposed approach works very well for the conductance peaks
associated with isolated resonances like the first peak
in Fig. \ref{GG1}. For the resonance (1,1)
the function $G_F + G_0$, red dashed line in Fig. \ref{GG1},
computed for $V_d=V_0^{(1,1)}$ for which $E_0^{(1,1)} \simeq E_F$,
provides an excellent approximation of the conductance on the peak.
The Fano parameter of the function $G_F$,
given in Tab. \ref{tabel},
corresponds to a slight asymmetry of the line shape.
This behaviour is expected for an isolated resonance
that interacts very weakly with the neighbor resonances and for which the
Fano type interference is negligible.

The profile of the peak (1,1) and the climb 
by $\pi$ of the phase $\varphi_{11}$ in the energy domain of this peak, 
Fig. \ref{GG1}, 
could suggest that an approach by means of a Lorentzian is actually enough
for describing isolated resonances. But the nonzero imaginary part
of the asymmetry parameter $q_F$ requires a detailed analysis of the phases
$\varphi_{11}$ and $\varphi_{F11}$,
Eqs. (\ref{phase}) and (\ref{Fano_phase}), respectively.
Around the resonance (1,1) the  energy dependence of 
$\varphi_{F11}$ is similar to that plotted in Fig. \ref{fano_g}
for $q_r=-50$ and $q_i=1.2$.
However, around the maximum $(1,1)$, 
where the Fano approximation 
for the total transmission $T(E)$ works well, only one of the two variations 
of $\pi$ in the transmission phase occurs. The second step in $\varphi_{F11}$ 
is outside the plotted interval. 
\begin{figure}[th]
\includegraphics*[width=2.75in]{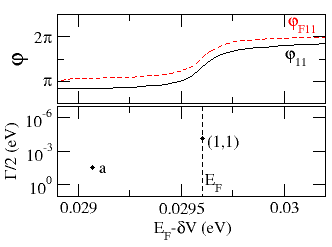}
\caption{(Color online)
Upper part: Transmission phase $\varphi_{11}$ (black full line)
around the resonance $(1,1)$ and its Fano approximation 
$\varphi_{F11}$ (dashed red line). The difference between the two
curves gives the constant phase $\varphi^{(0)}_{11}$.
Lower part: Poles and position
of the Fermi level in the complex energy plane for $V_d=V_0^{(1,1)}$.
}
\label{fi_max_1_1}
\end{figure}
As illustrated in Fig. \ref{fi_max_1_1}, 
for the resonance (1,1)
the Fano phase $\varphi_{F11}$ approximates excellently
the transmission phase $\varphi_{11}$;
The initial phase $\varphi_{11}^{(0)}$, Eq. (\ref{fi_a}), can not be 
determined, but it is not really relevant.

Around the resonance (1,1) the transmission phase $\varphi_{22}$
is practically constant, see Fig. \ref{GG1}, and this behaviour 
certifies also the odd parity of the resonant mode (1,1).
In contrast, for the resonance (2,2), which is also an isolated one,
the transmission phase $\varphi_{11}$ remains almost constant
in the resonance domain and $\varphi_{22}$ increases by $\pi$.
The slow variation of $\varphi_{11}$ around the resonance (2,2) is
determined by the influence of the mode (3,1) that is very broad.
Due to the symmetry reasons the modes (2,2) can not couple to
the neighbor modes (2,1) and (3,1) and 
the resonance (2,2) shows the typical features of an isolated resonance.

\subsection{Overlapping resonances}
\label{ov_res}

In the case of two overlapping resonances, $\bar{E}_{0 \lambda}$ 
and $\bar{E}_{0 \lambda'}$ with 
$\Gamma_{\lambda'} > \Gamma_{\lambda}$, around the Fermi energy,
each of them yields a resonant contribution to the conductance,
$G_{res}$ and $G'_{res}$, Eqs. (\ref{Gres}) and (\ref{Gresp}),
respectively.
The peak in the total transmission\cite{roxana10},
a superposition of $G_{res}$ and $G'_{res}$,
has an asymmetric line shape and a maximum value different from 1.
Similar to $G_{res}$, Eq. (\ref{GF}), the contribution
$G'_{res}$ of the second resonance 
can be also approximated as a Fano line
\begin{eqnarray}
G'_{res}(E_F-\delta V)
& \simeq &
G'_{1 \lambda}
\left| \frac{1}{v'_\lambda +i} + \frac{1}{q'_{F \lambda}} \right|^2
-G'_{2 \lambda} 
\nonumber \\
& = & G'_F(E_F-\delta V)
\label{GFp}
\end{eqnarray}
with $v'_\lambda = 2(E_F-\delta V-E_{0 \lambda'})/\Gamma_{\lambda'}$.
The complex asymmetry parameter $q'_{F \lambda}$ 
of this line is obtained from Eq. (\ref{qFF})
and the constants $G'_{1 \lambda}$ and $G'_{2 \lambda}$ 
from Eqs. (\ref{F1F}) and (\ref{F2F}), respectively.
In the case of only two overlapping resonances, 
the second background contribution 
to the conductance is approximatively a constant,
\begin{equation}
G'_{bg}(E_F-\delta V) \simeq G'_{bg}(E_F) = G'_0.
\label{G0p}
\end{equation} 

The broader resonance $\lambda'$ around the
resonance $\lambda$ influences not only the background term $G_{bg}$
in the conductance, Eq. (\ref{Ga}), but also the 
Fano asymmetry parameter $q_{F \lambda}$
associated with the first resonant term $G_{res}$, Eq. (\ref{GF}).
The function ${\cal{Z}}_{2 \lambda}$, Eq. (\ref{Z2}),
that enters $q_{F \lambda}$, Eq. (\ref{qFF}),
describes the interaction between resonances
and can be also decomposed in a resonant term singular 
at $E=\bar{E}_{0 \lambda'}$ and a background term,
\begin{equation}
{\cal{Z}}_{2\lambda}(E) 
= \frac{1}{{\cal{Z}}_{1 \lambda}(E)}
  \left[ \frac{\vec{\beta}_{1\lambda}^\dagger
               \cdot
               \vec{\beta}_{1\lambda'}
               \,
               \vec{\beta}_{2\lambda}^\dagger
               \cdot
               \vec{\beta}_{2\lambda'}}
               {E - E_{\lambda'}-\bar{\cal{E}}'_{\lambda}}
         +
         \vec{\beta}_{1\lambda}^\dagger 
         \si'_{\lambda}
         \vec{\beta}_{2\lambda}^*
   \right].
\label{Z22}
\end{equation}
The vector $\vec{\beta}_\lambda$ characterizes the resonance $\lambda$,
$\vec{\beta}_{\lambda'}$ the resonance $\lambda'$ and
$\si'_{\lambda}$ the other ones. The strength of the coupling 
between the two overlapping resonances is indicated by the
scalar products 
$\vec{\beta}_{1\lambda}^\dagger
 \cdot
 \vec{\beta}_{1\lambda'}$
and 
$\vec{\beta}_{2\lambda}^\dagger
 \cdot
 \vec{\beta}_{2\lambda'}$;
For resonant modes with the same symmetry the two vectors 
are approximatively parallel to each other,
while for different symmetries they are quasi-orthogonal to each other.
The resonant term in ${\cal{Z}}_{2 \lambda}$
is dominant only for the same parity. If this condition is fulfilled
$|1/q_{F \lambda}|$ becomes large and the transmission shows
a strong asymmetric line shape or a dip. Thus, by a favorable parity 
two overlapping resonances define two interfering pathways 
responsible for the Fano effect.
In the opposite situation, for resonant states $\lambda$ and $\lambda'$
with different parities,
the value of $|1/q_{F \lambda}|$ is quite small and the 
Fano line is almost symmetric. Therefore, the quantum interference 
of the two pathways plays a minimal role. However,
 the interference with other resonant pathways belonging to the 
background is not excluded.

If we perform further a drastic approximation in $1/q_{F \lambda}$,
neglecting the derivatives of the first order of
${\cal{Z}}_{1 \lambda}$ and 
$\bar{\cal{E}}_\lambda$, which are slowly varying functions of energy, 
we find
\begin{equation}
\frac{1}{q_{F \lambda}} 
\simeq -i \frac{i \Gamma_\lambda/2}
               {E_{0 \lambda}-E_{0 \lambda'}+i \Gamma_{\lambda'}/2}
          \left. \frac{\vec{\beta}_{1\lambda}^\dagger
                       \cdot
                       \vec{\beta}_{1\lambda'}
                       \,
                       \vec{\beta}_{2\lambda}^\dagger
                       \cdot
                       \vec{\beta}_{2\lambda'}}
                      {\vec{\beta}_{1\lambda}^\dagger
                       \cdot
                       \vec{\beta}_{1\lambda}
                       \,
                       \vec{\beta}_{2\lambda}^\dagger
                       \cdot
                       \vec{\beta}_{2\lambda}}
          \right|_{E=E_{0 \lambda}}.
\label{qFla_a}
\end{equation}
In the case of a favorable symmetry the second fraction is approximatively 1
and the relative position of the two overlapping resonances in the
complex energy plane determines the value of the asymmetry 
parameter $1/q_{F \lambda}$. For two resonances 
very close in energy, $E_{0 \lambda} \simeq E_{0 \lambda'}$,
the Fano parameter is a complex number with a large imaginary part
comparatively to the real one. This property of the asymmetry 
parameter is the fingerprint of an unitary dephasing in a quantum system
coupled to the environment\cite{brouwer01,srotter10}
and can be directly seen in the phase of the Fano function.
For different values of the Fano asymmetry parameter the phase shift 
is analyzed in Appendix \ref{fano_fct_c}.
Thus, based on a very simple model for the Fano asymmetry parameter,
we can associate dephasing effects to overlapping resonances 
that interact strongly due to the same symmetry of the resonant
modes.

Further we analyze different types of overlapping resonances 
and try to answer the question if the coupling between resonances
always induces a decoherent dephasing or not. Is the absence of the Fano
effect, i.e. a quasi symmetrical peak in the total transmission,
an unambiguous proof of the phase coherence?

\subsubsection{Weak interacting regime}
\label{weak}

\begin{figure*}[th]
\begin{center}
(a)
\noindent\includegraphics*[width=2.25in]{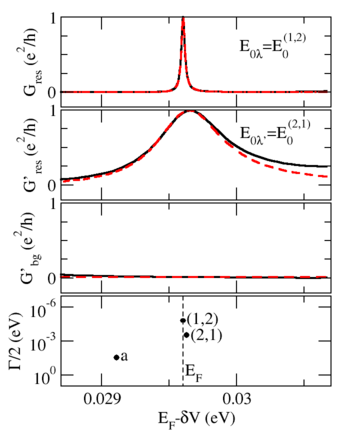}
(b)
\noindent\includegraphics*[width=2.25in]{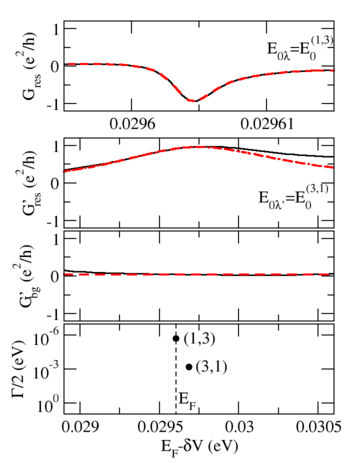}
\end{center}
\caption{(Color online)
Upper parts: Resonant contributions to the conductance
$G_{res}$ and $G'_{res}$ (solid black lines)
and the corresponding Fano lines $G_F$ and $G'_F$ (dashed red lines);
Middle parts: Background contribution to the conductance
$G'_{bg}$ (solid black lines) and its constant approximation 
$G'_0$ (dashed red lines).
Lower parts: Poles and position
of the Fermi level in the complex energy plane. The potential
energy in the dot region is constant, (a) $V_d=V_0^{(1,2)}$  
and (b) $V_d=V_0^{(1,3)}$.}
\label{2_ov_res}
\end{figure*}

In the weak coupling regime each resonance
contributes to the conductance with a Fano line
with a slight ($|1/q_F| \ll 1$)
up to an intermediate ($|1/q_F| \simeq 1$) asymmetry,
as plotted in Fig. \ref{2_ov_res}(a)
for the resonances (1,2) and (2,1).
The Fano parameters of these lines, 
see Tab. \ref{tabel}, reflect the
weak coupling of the resonances to each other and
to the neighbor ones\cite{roxana10}.
Because of its symmetry the resonant mode (1,2) is quasi isolated
and the corresponding Fano line shows practically a Breit-Wigner profile.
The parity of the resonant modes suppresses
in this case the Fano effect.
In contrast, the second mode, (2,1), interacts with the resonant mode (1,1)
as well as with the very broad resonance (3,1).
Its Fano line is slight asymmetric with $|1/q_F^{(2,1)}| > |1/q_F^{(1,2)}|$. 
Even in the case of a partial overlapping,
a favorable parity of the resonant modes allows for quantum interference
directly reflected by the asymmetry of the transmission peak.

\begin{figure*}[tbh]
\begin{center}
(a)
\noindent\includegraphics*[width=2.25in]{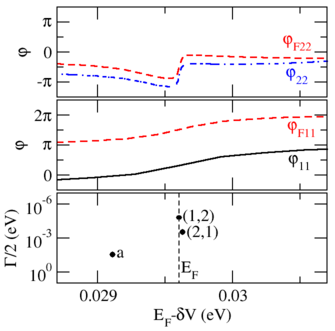}
(b)
\noindent\includegraphics*[width=2.25in]{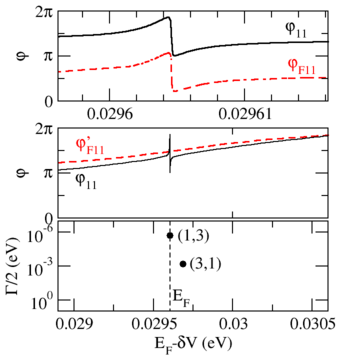}
\end{center}
\caption{(Color online)
Upper parts: Transmission phases $\varphi_{nn}$, (solid black lines 
for $n=1$ and dot-dashed blue line for $n=2$) and the Fano phases 
$\varphi_{Fnn}$ (red dashed lines) for the resonances (1,2) and (2,1)
in (a) and (1,3) and (3,1) in (b).
Lower parts: Poles and position
of the Fermi level in the complex energy plane. The potential
energy in the dot region is constant, (a) $V_d=V_0^{(1,2)}$ 
and (b) $V_d=V_0^{(1,3)}$.}
\label{2_ov_res_fi}
\end{figure*}

Although the first Fano line is quasi-symmetric,
the interaction between the resonances (1,2) and (2,1)
is present and leads to their separation
on the imaginary axis\cite{roxana10}.
A second manifestation of this interaction
is the decoherent dephasing of the resonant mode (1,2),
directly observed in the transmission phase $\varphi_{22}$ plotted 
in Fig. \ref{2_ov_res_fi}(a).
This phase presents a drop of about $\pi/2$ followed by a climb of $\pi$
through the  resonance (1,2), while the phase $\varphi_{11}$ increases
monotonically by $\pi$ through the  resonance (2,1).
To prove that the interaction of the overlapping resonances
is responsible for dephasing, we analyze the imaginary part
of the Fano asymmetry parameter associated
with the transmission probability $T_{22}$, see Tab. \ref{tabel}.
The value of this parameter, $|1/q_{F22}|\simeq 0.04$,
corresponds to a quasi symmetric profile in $T_{22}$,
similar to $G_{res}$ in Fig. \ref{2_ov_res}(a),  but the imaginary part 
of $q_{F22}$ is larger than the real one. According to the analysis in
Appendix \ref{fano_fct_c} the corresponding Fano phase $\varphi_{F22}$
shows a dip and not a monotonically increasing by $\pi$ as could be 
expected for a quasi Breit-Wigner line.
The Fano phase $q_{F22}$ describes accurately
the phase evolution through the resonance (1,2)
as illustrated in Fig. \ref{2_ov_res_fi}(a). 
Also in the case of the resonance (2,1) the phase defined in the 
frame of the Fano approximation, $\varphi_{F11}$, describes 
very well the transmission phase $\varphi_{11}$; the 
associated asymmetry parameter is given in Tab. \ref{tabel}. 

Here we have to remark, that, for some resonances,
there is a difference between the Fano parameter $q_F$ 
associated with $T$ and the parameters $q_{F11}$ or $q_{F22}$ 
associated with $T_{11}$ or $T_{22}$, respectively.
The total transmission $T$ describes a global effect, containing 
contributions from all scattering channels while $T_{11}$ and $T_{22}$ 
correspond only to the transmission between a pair of channels.
As reported in Ref. \onlinecite{ando04},
the global Fano parameter is given as a 
linear combination of the asymmetry parameters for 
each pair of scattering channels. Thus, the corresponding profiles in the
transmission differ slightly from channel to channel and 
also from the total transmission.

A similar phase shift across the second peak in the conductance 
is observed experimentally by Kalish et al.\cite{kalish05}. 
They reported a dip in the transmission phase that always occurs
in front of the second conductance maximum, i.e. before the entry of the 
second electron for the setup in that experiment. The drop of about
$\pi/2$ is followed by a phase antilapse\cite{oreg02} and 
these phase variations survive the 
changes in the system parameters.
The experimental findings are especially interesting if we consider that 
the quantum dot used by Kalish et al. \cite{kalish05}
and the dot modelled by us 
have similar geometries but different electron densities. 
We can conclude that the phase antilapse associated with the overlapping 
resonances (1,2) and (2,1) is a general finding and it
occurs independently of the quantum system parameters.
Responsible for this  phase evolution can be only
the dephasing process induced by the interaction of the 
two overlapping resonances. 
While this interaction leads to the quantum interference and 
consequently to asymmetric Fano line in the total transmission 
only in the case of a favorable parity of the resonant modes,
the dephasing is present even for the weak coupling regime of the
overlapping resonances.

\begin{figure}[tbh]
\begin{center}
\noindent\includegraphics*[width=2.25in]{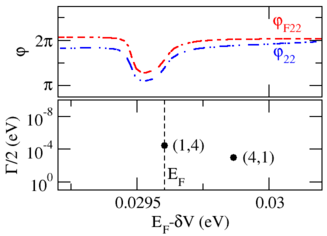}
\end{center}
\caption{(Color online)
Upper part: Transmission phase $\varphi_{22}$ (dot-dashed blue line) 
and the Fano phase $\varphi_{F22}$ (red dashed line).
Lower parts: Poles and position
of the Fermi level in the complex energy plane. The potential
energy in the dot region is constant, $V_d=V_0^{(1,4)}$. 
}
\label{1_4_fi}
\end{figure}

A second antilapse in the transmission phase, also a so-called 
nonuniversal one\cite{landman08}, occurs in the energy
domain of the resonance (1,4), as plotted in Fig. \ref{1_4_fi}. 
For this resonance the Fano asymmetry parameter $q_{F22}$, Tab. \ref{tabel},
indicates a slight asymmetry of the transmission line $T_{22}$ 
and an approximative symmetrical dip of about 
$3\pi/4$ in the transmission phase $\varphi_{F22}$. 
Also in this case the decoherent dephasing of the resonant 
modes (1,4) and (4,1) occurs as an effect of the  interaction between
resonances, accompanied by the separation of the two 
resonances on the imaginary axis\cite{roxana10}.

\begin{figure}[htb]
\begin{center}
\noindent\includegraphics*[width=2.25in]{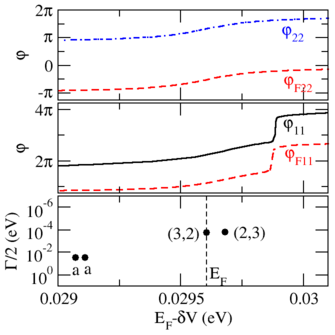}
\end{center}
\caption{(Color online)
Upper parts: Transmission phases $\varphi_{22}$ (dot-dashed blue line) 
and $\varphi_{11}$ (solid black line) and the Fano phases 
$\varphi_{F11}$ and $\varphi_{F22}$ (red dashed lines).
Lower part: Poles and position
of the Fermi level in the complex energy plane. The potential
energy in the dot region is constant, $V_d=V_0^{(3,2)}$. 
}
\label{2_3_fi}
\end{figure}

In contrast, the weakly interacting resonances (2,3) and (3,2)
yield an asymmetric peak in conductance, but  
no phase lapse or antilapse in the energy domain of this peak, 
see Fig. \ref{GG2}.
The reason is the relative large width of the resonance (3,2)
that leads to a broad maximum in the transmission phase from which only
a part is included in the energy domain of the transmission peak,
as plotted in Fig. \ref{2_3_fi}. The Fano approximation 
for the total transmission and for the transmission phases
works also in this case very well.

\subsubsection{Strong interacting regime}
\label{strong}

Opposite to the weak interaction regime, the strong interaction 
between overlapping resonances requires the same parity in the 
lateral direction of the corresponding resonant modes.
As analyzed in Ref. \onlinecite{roxana10},
the strong resonance coupling leads to
the hybridization of the resonant modes
and to a strong repulsion 
of the resonances  in the complex energy plane,
i.e. a {\it resonance trapping}\cite{rotter09,mueller09}.
In turn,  strong localized resonant states, i.e. quasi-bound states, occur
accompanied by broad ones, all of them with probability
distribution densities in the dot region without correspondent 
between the eigenstates of an isolated dot\cite{roxana10}. 
The  conductance peak
associated with two strong interacting resonances 
is generally 
the superposition of a thin dip and a broad slight asymmetric maximum,
as plotted in Fig. \ref{2_ov_res}(b) for 
the resonances (1,3) and (3,1).
The dip in conductance corresponding to the resonance (1,3)
is characterized by a large absolute value of 
$|1/q_F|$, Tab. \ref{tabel}, reflecting  the presence of 
the quantum interference.
The real energies of the resonances (1,3) and (3,1)
are very close and they define two interfering pathways
for the transmission through the quantum system. The first path corresponds
to the narrow resonance, while the second one is associated with a 
continuum of states in the energy domain of the second broad resonance.
For this reason we can speak about a Fano type interference in open
quantum systems strongly coupled to the environment.

A second effect associated with interacting resonances is the 
nondissipative dephasing induced on the resonant modes. 
As illustrated in the upper part of Fig. \ref{2_ov_res_fi}(b),
in the energy domain of the resonance (1,3) 
the transmission phase $\varphi_{11}$ 
increases by $\sim \pi/2$, shows after that a phase lapse of $-\pi$
and increases again to the initial value. 
The phase variation of $\varphi_{11}$ is excellently
described by the Fano phase. 
The imaginary part of the asymmetry parameter 
$q_{F11}$, Tab. \ref{tabel}, exceeds the real one and 
this peculiarity indicates the presence of the dephasing process 
in the case of interacting resonances, irrespective of the 
coupling strength between them. The value of the complex asymmetry parameter 
$q_{F11}$ for the resonance (1,3) corresponds to a lapse 
in the phase $\varphi_{F11}$ with a mesoscopic character. In contrast to the
universal phase lapses, the nonuniversal ones depend on the dot occupancy. 

The broad resonance (3,1) yields a supplementary
variation of the phase $\varphi_{11}$ by $\pi$ within a much larger energy
interval than the energy domain of the resonance (1,3). 
The phase $\varphi'_{F11}$, Fig. \ref{2_ov_res_fi}(b) middle part, 
approximates also well the second slowly variation in $\varphi_{11}$.
Opposite to the case of weakly interacting resonances,
in the strong coupling regime both resonances determine variations
of the same transmission phase. Thus, for the resonances (1,3) and (3,1)
the transmission coefficients between odd scattering channels show  
phase variations, while in the case of the resonances (2,4) and (4,2) only 
for the even channels the phase is shifted through the 
overlapping resonances.

\begin{figure}[tbh]
\begin{center}
\noindent\includegraphics*[width=2.25in]{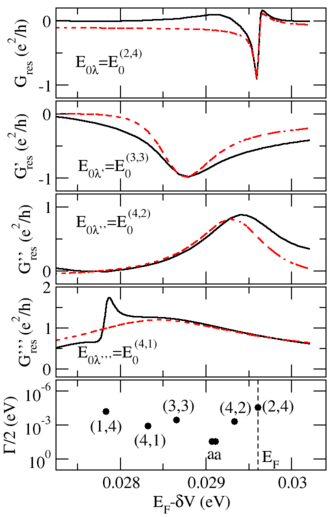}
\end{center}
\caption{(Color online)
Upper parts: Resonant contributions to the conductance
$G_{res}$, $G'_{res}$, $G''_{res}$ and $G'''_{res}$ (solid black lines)
and the corresponding Fano lines $G_F$, $G'_F$, $G''_F$ and $G'''_F$
(dashed red lines); The background contribution is
$G'''_0 = 0.36 \, e^2/h$.
Lower part: Poles and position
of the Fermi level in the complex energy plane for $V_d=V_0^{(2,4)}$.
}
\label{2_4_fano}
\end{figure}

In Ref. \onlinecite{roxana10} we have already shown, that the 
overlapping resonances (2,4) and (4,2) interact strongly to each other
and they are associated with typical hybrid resonant modes. But each effort
to decompose the conductance in only two resonant contributions and a 
background was unavailing. Increasing the dot occupancy, increases also
the resonant level density around the Fermi energy and there are more than two
resonances that overlap and interact. For $V_d=V_0^{(2,4)}$
the four resonances that overlap are plotted 
in the lower part of Fig. \ref{2_4_fano}. 
The first contribution 
to the conductance is given by the resonance (2,4). This is 
the thinest one, followed by the resonances (3,3), (4,2) and (4,1)
with $\Gamma^{(2,4)} < \Gamma^{(3,3)} < \Gamma^{(4,2)} < \Gamma^{(4,1)}$. 
This arrangement is determined by the existence of
two pairs of strongly interacting resonances: (2,4) and (4,2) and 
(3,3) and (4,1), respectively. Corresponding to the
strong interactions the first two contribution to the conductance,
$G_{res}$ and $G'_{res}$, are minima described by Fano lines.
They are plotted in the upper parts of Fig. \ref{2_4_fano}.
Comparatively to the pair of strong interacting resonances (1,3) and (3,1), 
Fig. \ref{2_ov_res}(b), in this case
the first dip in conductance is broader because the strong 
interaction between the resonances (2,4) and 
(4,2) is perturbed by their weak coupling to the resonance (3,3).
The other two contributions to
the conductance, $G''_{res}$ and $G'''_{res}$ in Fig. \ref{2_4_fano},
are broad maxima described also by Fano lines. 
The conductance is obtained in this case as a superposition of four
Fano lines and a constant background.
The asymmetry parameters of the Fano lines are given in Tab. \ref{tabel}.
As illustrated in Fig. \ref{GG2},
the Fano approximation is satisfactory even in the case of 
four overlapping resonances
and validates the description of the open quantum
systems in terms of resonances. Due to the mutual interactions,  
the overlapping resonances are spread out into the complex energy plane
so that there exists always a hierarchy of the resonances and the 
decomposition in successive Fano lines  can be done. The asymmetry parameter
of each line gives information about the coupling strength of the
associated resonance to the other ones and, implicitly,
about the presence or not of the quantum interference. 

In the case of the four overlapping resonances analyzed above,
the transmission phase $\varphi_{11}$ associated with odd channels
as well as the phase $\varphi_{22}$ associated with the even ones
show variations around the Fermi energy.
\begin{figure}[htb]
\begin{center}
\noindent\includegraphics*[width=2.25in]{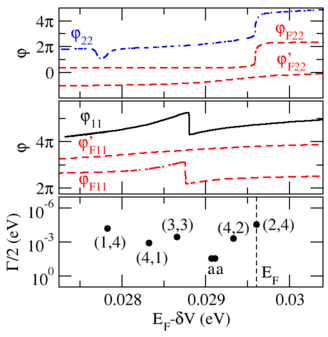}
\end{center}
\caption{(Color online)
Upper parts: Transmission phases $\varphi_{22}$ (dot-dashed blue line)
and $\varphi_{11}$ (solid black line) 
and the Fano phases $\varphi_{F11}$, $\varphi_{F22}$, $\varphi'_{F22}$
and $\varphi'_{F11}$ (dashed red lines) associated with the
resonances (2,4), (3,3), (4,2) and (4,1), respectively.
Lower part: Poles and position
of the Fermi level in the complex energy plane for $V_d=V_0^{(2,4)}$.
}
\label{2_4_fi}
\end{figure}
The Fano phase $\varphi_{F22}$ determined by the asymmetry
parameter $q_{F22}$ corresponding to the resonance (2,4), Tab. \ref{tabel}, 
approximates very well the abrupt jump of $2\pi$ in the
transmission phase $\varphi_{22}$.
The second slow variation of $\pi$
in $\varphi_{22}$, superimposed on the jump, is determined
by the resonance (4,2). This shift is also well described 
in the frame of the Fano approximation, as illustrated in the 
upper part of Fig. \ref{2_4_fi}.   
The strongly interacting resonances (3,3) and (4,1)  determine 
in $\varphi_{11}$  a nonuniversal 
phase lapse superimposed also on a slowly variation of $\pi$  associated
with the broader resonance, see middle part of Fig. \ref{2_4_fi}.
The Fano asymmetry parameters corresponding to the 
resonances (3,3) and (4,1) 
are given in Tab. \ref{tabel}.
The Fano phases $\varphi_{F11}$ and $\varphi'_{F11}$
associated with these parameters approximate  well the phase shift of 
$\varphi_{11}$.

\section{Conclusions}

In this paper we study the transmission probabilities and phases 
of noninteracting spinless electrons
transfered through an open quantum dot strongly coupled 
to the conducting leads. 
The mechanism that governs the transport properties in the strong
coupling regime is the interaction between overlapping resonances.
Due the coupling of the quantum system to the environment, 
a degenerate or quasi-degenerate energy level evolves to 
two overlapping resonances that interact. 
%They can not evolve independently,
%they influence each other.
As a consequence of the interaction the two resonances are separated
in the complex energy plane
and the corresponding resonant modes are dephased with respect
to each other.  
The induced dephasing is the necessary condition for the 
%appearance of the 
quantum interference, but this condition  is not sufficiently. 
The Fano effect requires supplementary a favorable parity of the
resonant modes.

Using a model based on the scattering theory  
and the R-matrix formalism, we propose a successive decomposition 
of the total transmission in contributions corresponding to each 
resonance of the overlapping ones. Each contribution is approximated
as a Fano line with a complex asymmetry parameter.
The imaginary part of the Fano parameter 
gives information about the dephasing degree of the
corresponding resonant modes, while its modulus
characterizes the quantum interference with the other resonant modes. 

The effects of the interaction between resonances can be directly seen 
in the measurable quantities conductance and transmission phases
through the open quantum dot. In the frame of the Fano approximation 
we provide a very good description of the dips in conductance 
superimposed on broad maxima and accompanied by nonuniversal
lapses in the transmission phases. We analyze mainly the mesoscopic 
regime characterized by a low occupancy of the quantum dot
for which only two resonances interact. We associate the strong coupling 
between resonances with the Fano effect, while in the
weak coupling regime only the shift of the transmission phases
indicates the presence of their interaction.  
By increasing further the dot occupancy a set of many overlapping resonances
lies around the Fermi energy and participates to the transport.
The resonances interact with each other and
determine complex line shapes in conductance and many transmission 
phase lapses.
We can speak about the crossover regime between the mesoscopic and the
universal one\cite{kalish05}. 
For a proper characterization of this regime we have also to consider 
the Coulomb interaction that becomes nonnegligible by 
increasing the number of the electrons within the dot region.
This subject will be studied in the future.

\appendix

\section{Laurent expansion of the resonant term around the pole}
\label{laurent}

We consider the general form of the resonant term in transmission
as a complex function 
\begin{equation}
f(E) = \frac{2 i z_{1 \lambda}(E)}
            {E - E_\lambda -\bar{e}_\lambda(E)}
        -z_{2 \lambda}(E),
\label{fct_f}
\end{equation}
with a simple pole at $E=\bar{E}_{0 \lambda}$ and with 
$\bar{e}_\lambda$, $z_{1 \lambda}$ and $z_{2 \lambda}$ 
slowly varying functions in the energy domain of the resonance $\lambda$.

To obtain the line shape of the total transmission, $F(E)=|f(E)|^2$,
around the considered resonance, we extend the function $f(E)$
by analytic continuation in the lower part of the complex energy plane 
and employ a formal expansion of it in a Laurent series around the pole.
The slow variation with the energy of the functions
$\bar{e}_\lambda$, $z_{1 \lambda}$
and $z_{2 \lambda}$  around the resonance $\lambda$ 
allows to neglect  their derivatives at the resonance energy 
up to the second order\cite{roxana01}. 
Thus, for energy around $E_{0 \lambda}$, the function $F(E)$ 
can be approximated as a Fano line
\begin{equation}
F(E) \simeq F_{1 \lambda} \left| \frac{1}{e_\lambda +i} 
                                +\frac{1}{q_{F \lambda}} 
                          \right|^2
           -F_{2 \lambda},
\label{FF}
\end{equation}
where 
$e_\lambda = 2(E-E_{0 \lambda})/\Gamma_\lambda$.
The Fano asymmetry parameter $q_{F\lambda}$ is a complex number defined as
\begin{equation}
\frac{1}{q_{F \lambda}}
 = i \left.
     \left[ \frac{d z_{1 \lambda}}{d E}
           -i \frac{z_{2 \lambda}}{2}
            \left(1 - \frac{d \bar{e}_\lambda}{d E}
            \right)
      \right]
      \left[ \frac{d z_{1 \lambda}}{d E}
            +i \frac{z_{1 \lambda}}{\Gamma_\lambda/2}
      \right]^{-1}
     \right|_{E=E_{0 \lambda}}
\label{qFF}
\end{equation}
and the constants $F_{1 \lambda}$ and $F_{2 \lambda}$ have the expressions
\begin{equation}
F_{1 \lambda} = \left.
                \left| \frac{d z_{1 \lambda}}{d E}
                      +i \frac{z_{1 \lambda}}
                              {\Gamma_\lambda/2}
                \right|^2
                \left|1 - \frac{d \bar{e}_\lambda}{d E}
                \right|^{-2}
                \right|_{E=E_{0 \lambda}}
\label{F1F}
\end{equation}
and
\begin{equation}
F_{2 \lambda} = \left| z_{2 \lambda}(\bar{E}_{0 \lambda}) \right|^2,
\label{F2F}
\end{equation}
respectively.

We have already used this approach to determine the 
conductance around an isolated resonance in the case of a separable
confinement potential in the dot region,
i.e. for an effective 1D system\cite{roxana01}.
But for the 2D case,
the complex function $\bar{\cal{E}}_\lambda(E)$, 
which assures the analyticity of the resonant term, Eq. (\ref{Stilde3}), 
becomes a product of two infinite vectors $\vec{\alpha}_\lambda$
and $\vec{\alpha}_\lambda^T$ with the infinite matrix 
$(1 + i \mathbf{\Omega_\lambda})^{-1}$. 
For the numerical calculations we have to cut
$\vec{\alpha}_\lambda$ and $\mathbf{\Omega_\lambda}$.
The confinement potential of the quantum dot analyzed here 
has no attractive character\cite{simon76} 
and the influence of the evanescent channels can be neglected
in this case\cite{racec09}.
As usually done in the Landauer-B\"uttiker 
formalism\cite{buettiker85,buettiker86},
we have considered 
 only the contributions of the open channels, 
i.e. a $2N_F$ vector $\vec{\alpha}_\lambda$ 
and a $2N_F \times 2 N_F$ $\mathbf{\Omega_\lambda}$  matrix,
where $N_F=N_1(E_F)=N_2(E_F)$.

\section{Fano functions}
\label{fano_fct_c}

The complex Fano function is defined as
\begin{equation}
f_c(e) = \frac{1}{e+i} + \frac{1}{q_F} 
       = \frac{1}{|q_F|} \, \sqrt{f(e)} \, e^{i\varphi_F},
\label{fc}
\end{equation}
where $q_F=(q_r,q_i)$ is a nonzero complex number 
usually called asymmetry parameter. 
This parameter is
responsible for the asymmetry of the real Fano function\cite{fano,fano_book} 
\begin{equation}
f(e)=\frac{|e+q|^2}{e^2+1},
\end{equation}  
with $q=q_F+i$.
For each value of the parameter $q_F$ the function $f(e)$
has a maximum and a minimum,
$$
M = 1 + \frac{q_r}{e_M} = f(e_M)
\quad
\text{and}
\quad
m = 1 + \frac{q_r}{e_m} = f(e_m)
$$
respectively,
for 
\begin{equation}
e_{M,m} = -\frac{1}{2q_r}
           \left[ 2q_i+|q_F|^2 \mp \sqrt{(2q_i+|q_F|^2)^2+4q_r^2}
           \right].
\end{equation}
In the limit $|1/q_F| \rightarrow 0$ the function $f(e)$ reduces to the
Breit-Wigner form with a maximum at the origin, $e_M \rightarrow 0$, 
and a minimum at infinity, $e_m \rightarrow \infty$. 

\begin{figure*}[tb]
\begin{center}
\includegraphics[width=6.5in]{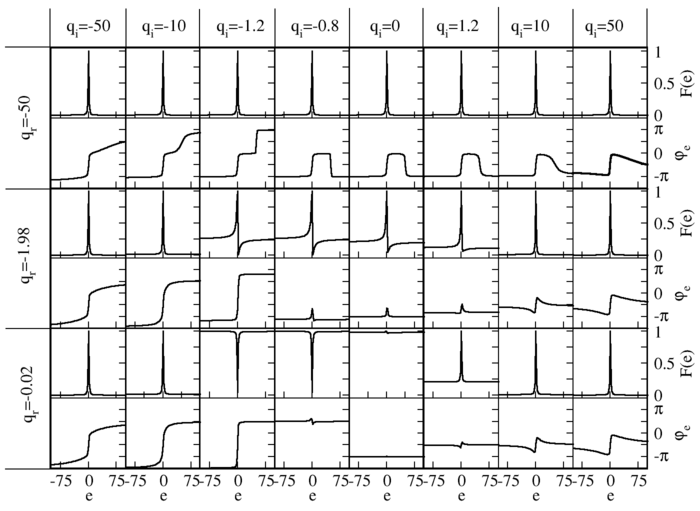}
\caption{
Fano function $F(e)=f(e)/M$ and the phase $\varphi_F$ of the complex 
Fano function, Eq. (\ref{fc}), for different values of the Fano asymmetry 
parameter $q_F=(q_r,q_i)$. }
\label{fano_g}
\end{center}
\end{figure*}

The phase of the complex Fano function 
is determined from the equation
\begin{equation}
\tan(\varphi_F) = -\frac{q_i (e^2 + 1) + |q_F|^2}
                        {q_r (e^2 + 1) + |q_F|^2 e}
\label{tan}
\end{equation}
in which $q_F$ enters as a parameter.
As plotted in Fig. \ref{fano_g}, the phase dependence on $e$
varies strongly with  the value of the Fano asymmetry parameter.
Further we provide a systematic analysis of the 
function $\tan(\varphi_F)$ in order to predict the phase evolution 
in the domain of a Fano line for a given value of $q_F$.
From symmetry reasons 
$\tan[\varphi_F(e;q_r,q_i)]=-\tan[\varphi_F(-e;-q_r,q_i)]$
and we have plotted only the curves for $q_r < 0$.

According to Eq. (\ref{tan}), 
$\lim_{e \rightarrow \pm \infty} \tan(\varphi_F) 
= -q_i/q_r =\tan(\varphi_\infty)$
and, consequently, the phase $\varphi_F$ has equal values 
at $e \rightarrow \pm \infty$
or shows a global variation of $2 m \pi$, $m \ge 1$ along the $e$ axis.
Because the function $\tan(\varphi_F)$ has only two zeros 
at $e_{z\pm} = \pm \sqrt{-|q_F|^2/q_i-1}$
for
\begin{equation}
%q \in \{q_i < 0, \,\,q_r^2+(q_i+1/2)^2 > 1/4 \},
q_F \in \{(q_r,q_i) \in {\bf C}, \, q_r^2+(q_i+1/2)^2 > 1/4 \},
\end{equation}
and only two poles at
$e_{p\pm} = -(|q_F|^2 \mp \sqrt{|q_F|^4-4q_r^2})/2q_r$
for
\begin{equation}
q_F \in \{(q_r,q_i) \in {\bf C}, \, (1-|q_r|)^2+q_i^2 > 1 \}
\end{equation}
the phase shift can not exceed $2 \pi$.
To differentiate further between the cases that correspond to a 
global variation of $2\pi$ in $\varphi_F$ and the cases
for which 
$\lim_{e \rightarrow -\infty}(\varphi_F)
 \simeq \lim_{e \rightarrow \infty}(\varphi_e)$,
it is necessary to analyze the monotonicity of $\tan(\varphi_F)$.
For 
\begin{equation}
q_F \in \{(q_r,q_i) \in {\bf C}, \, q_i > -1 \},
\end{equation}
the first derivative of $\tan(\varphi_F)$ has two zeros
at $e_{m\pm} = \left[ q_r \mp \sqrt{(1+q_i)|q_F|^2}\right]/q_i$
and they correspond to a minimum and a maximum of this function,
$\tan[\varphi_F(e_{m\pm})]
= 2 q_r e_{m\pm} \tan(\varphi_\infty) /(2 q_r e_{m\pm}+|q_F|^2)$.
The phase variation between the maximum and the minimum values, 
Fig.\ref{fano_g}, depends strongly on  $q_F$, but 
it can not exceed $\pi$. The limit value is obtained for $|q_F| \gg 1$.
In the case $q_i \ll |q_r|$, $|q_r| \gg 1$ the phase increases 
from $0$ to $\pi$, has a plateau and decreases approximatively to $0$, 
while in the case $|q_r| \ll q_i$, $q_i \gg 1$ the phase decreases 
from  $-\pi/2$ to $-\pi$, increases after that to $0$ and decreases 
further to $-\pi/2$. For smaller values of the Fano parameter
for which the asymmetry of the Fano line is strong, $|q_F| \le 1$, 
the phase does not vary strongly.

Opposite to this situation is the phase variation for $q_i < -1$. 
In this case $d \tan(\varphi_F)/de$ has no zeros and, in turn,
$\tan(\varphi_F)$ increases monotonically with $e$. The phase shows
a global variation of $2 \pi$. Depending on the values of the
Fano asymmetry parameter $|q_F|$, there is a jump of $2 \pi$ or two steps
of $\pi$ centered on $e_{p+}$ and $e_{p-}$. Actually the first case is a 
limit of the second one for a very small distance between 
$e_{p+}$ and $e_{p-}$. 
  
The systematic analysis of the complex Fano function, Eq. (\ref{fc}),
has shown that the phase of this function 
changes strongly by relative small variation of the  
real or the imaginary part of the asymmetry parameter.
Even in the case of quasi-symmetrical Fano lines, $|1/q_F| \ll 1$,
the quotient $q_i/q_r$ and the sign of $q_i$ yield 
a large variety of profiles in the phase evolution.

\bibliography{fano2}

\end{document}